\documentstyle[12pt]{article}

\begin{document}

\title{
\bf  DUALITY AND SCALING IN QUANTUM MECHANICS}               
 
\author{
{\bf Dhurjati Prasad Datta}\\                        
\normalsize Department of Mathematics \\                      
North Eastern Regional Institute of Science And Technology \\     
Itanagar-791109,Arunachal Pradesh, INDIA\thanks{Permanent address}\\              
\noindent and \\                                
Inter-University Centre for Astronomy and Astrophysics\\        
Ganeshkhind, Pune-411007, INDIA \\ \\                  
email:dpd@nerist.ernet.in}                

\date{}

\maketitle

\begin{abstract} 
  The nonadiabatic geometric phase  in  a  time  dependent  quantum 
evolution is shown to provide an intrinsic concept of time having dual 
properties relative to the external time.A nontrivial extension of the 
ordinary quantum mechanics is thus obtained with  interesting  scaling 
laws.A fractal like structure in time is thus revealed.
\end{abstract}

\newpage

     A generic property of a time dependent quantal evolution is  that 
the quantal system picks a nonzero  geometric  phase[1],  besides  the 
ordinary dynamical phase indicating the flow of time.The existence  of 
a geometric phase in turn indicates the  presence  of  a  small  scale 
motion  in  the  system[2,3],whose  origin  could  be  traced  to  the 
geometry(curvature)  of  the  projective  Hilbert  space.The  physical 
implications of this quantal geometric motion are  being  investigated 
in a series of papers  recently[3].The  motivation  of  these  studies 
comes mainly from quantum gravity(cosmology).A well known problem   in 
the canonical quantum gravity is the absence of an a  priori  external 
concept of time at the quantum level.The possible  emergence  of  time 
intrinsically from the dynamics of an interacting system  is  of  much 
relevance here.Indeed it is shown[3],in particular, that  an  internal 
concept of  time  could  be  realized  in  an  interacting  system  in 
connection with the nonadiabatic geometric phase picked by a pertinent 
quantal  state.Although  derived  in  the  framework  of   a   quantum 
cosmological  Wheeler-Dewitt  (WD)  equation,this  and  other  related 
results, however, turn out  to be more universal  in  character.It  is 
therefore  of  much  interest  to  investigate  analogous  results  in 
ordinary quantum mechanics.In this  letter,we  do  precisely  this  by 
reporting some preliminary,but novel results in this direction.

     The mean energy in  a  quantal  state  defines  a  scale  of  the 
external  (Newtonian)  time,which  in  turn  is  assumed  usually   to 
parametrize the path of the evolving state  in  the  relevant  Hilbert 
space.The actual evolution of the quantal state  however  consists  of 
two independent components: the dominant, purely dynamical,  evolution
in connection with the mean energy of the state,  accompanied  with  a 
small geometric  component  appearing  as  fluctuations  on  the  mean 
evolution.Treating the Hilbert space as a U(1) principal  bundle  over 
the projective space of rays,the  mean  evolution  could  in  turn  be 
represented as a pure vertical motion along the fibre corresponding to 
the state.This dominant vertical component in  the  quantal  evolution 
thus induces a change of phase in the state:the  so  called  dynamical 
phase.The (nonadiabatic) geometric  phase,  on  the  otherhand,  is  a 
consequence of the small scale geometric  motion  in  the  state.  The 
inherent geometric nature of this phase is captured by the  associated 
parallel transport law which not only states how the  state  is  being 
parallely transported along a horizontal curve in the projective space 
but   also   provides   a   natural   framework   for    its    actual 
computation[1,2].However,there is an equivalent  derivation[4]  which 
treats the geometric phase instead as a dynamical one, thus  realizing 
it as a correction term in the total phase.In the following  we  shall 
make use of this later approach in  our  discussion  of  the  internal 
time.

     As remarked already,the nonadiabatic geometric phase  is  related 
to  the   quantal   fluctuations   in   the   state   in   a   natural 
way.Consequently,the geometric contribution in the vertical motion  in 
the state is, in fact, due to these irreducible fluctuations.In  fact, 
we shall show that an intrinsic description of the  quantal  evolution 
could  be  made  meaningful  in  association  with  the   nonadiabatic 
geometric phase and hence the fluctuations  in  the  state.\footnote 
{ The adiabatic phase,on the contrary, is nonzero only for a parameter 
space with nontrivial geometry( topology).Although an internal concept 
of time could even be attached to a  nontrivial  adiabatic  phase,this 
would only be a special feature of the specific parameter space chosen 
to  describe  the  quantal  adiabatic  motion.The  internal  time   in 
connection with a nonadiabatic phase, on the other hand,  is  uniquely 
defined by the Fubini-study metric in the projective space  and  hence 
has a universal character(see the main text for details.} The  small 
vertical displacement of the state due to the nonadiabatic phase could 
indeed be interpreted locally (in the neighbourhood of a given instant 
of the external time) as  a  (dominant)  dynamical  evolution  in  the 
relatively small scale of the internal time variable,having  conjugate 
relationship  with   the   root   mean   square   energy   fluctuation 
(uncertainty).The term 'internal' qualifies the fact  that  it  is  an 
intrinsic gauge invariant geometrical object constructed  out  of  the 
quantal state itself (eg.,the Fubini-Study metric  in  the  projective 
space) without any reference  to  an  external  concept[5].In  quantum 
cosmology,this internal time variable, after a proper  rescaling,could 
be identified by an internal observer as the observable  time  in  the 
universe,reasonably well-defined for  the  description  of  the  local 
physics[3].Note that the fluctuations in the original state  not  only 
will appear relatively large O(1) with respect  to  this  small  scale 
internal time,but an (internal) observer equipped with  this  internal 
time and the associated rescaled  Schrodinger  equation  (cf.,eq.(11)) 
even identifies a 'large' (external) fluctuation as the mean energy in 
the pertinent internal state.Because  of  the  irreducibility  of  the 
quantal fluctuations,the above process could be iterated ad infinitum, 
thus establishing self-similar fluctuations at all (time) scales in  a 
non-stationary quantal evolution.

     Using the technique of the renormalization group (RG) as  in  the 
statistical mechanical models near a critical point,we  explore  these 
self-similar quantal fluctuations in  the  extended  framework  of  an 
internal time.The RG critical point corresponds here to the short time 
(large fluctuation) limit of the quantal evolution.In that  limit  the 
external  and  internal  time  variables  are  found  to   be   dually 
related,reminiscent     of     the     recently     studied     string 
dualities[6].Further,  the  nontrivial  scaling  of  the   two   point 
time-time correlator is interpreted as a signature of  the  fractality 
of time.We also explore some physical (experimental)  implications  of 
the fractal time.

   Let us consider a quantal evolution of a state $|\psi(t)>$ given by 
the Schrodinger equation ($\hbar =1$)
$$
{i{d \over {dt}}|\psi> = H |\psi>} \eqno(1)
$$ 

\noindent where $H$ denotes the  Hermitian  Hamiltonian  operator  and  
$t$  is  the 
external time. For definiteness,we assume that  the  potential  energy 
function is bounded from below.Let $H$ denote the Hilbert space  of  the 
normalized states:$<\psi|\psi>=1$.Then $H$ is the U(1) principal bundle  
on the 
projective space $\cal P$ of rays. Recall[7] that a nonstationary
isolated 
state moves around $\cal P$ provided the corresponding uncertainty $\Delta E$
in the 
energy:$\Delta E =\sqrt{<\psi| (H-E)^2|\psi>}, E= <\psi|H\psi>$,is  
nonzero. Further, the 
characteristic time scale of the dynamics is fixed by $t\approx 1/E$ with  
the 
auxiliary condition that $\Delta E \rightarrow 0$ in the adiabatic limit  
$t\rightarrow t_0$,which we 
assume as the initial condition. We restrict the discussion to the 
evolution close to $t\rightarrow t_0$, when the nonadiabatic 
corrections become 
important. More general initial conditions will be treated elsewhere.

     In the presence of a geometric phase (nonzero for a nonstationary 
state) the time derivative in the  Schrodinger  equation  (1)  can  be 
interpreted as a total derivative 
$$
{{D\over dt} = {\partial \over {\partial}t} + {d \over dt}
= {\partial \over {\partial}t} + 
{dq_i \over dt}{\partial \over {\partial}q_i}}\eqno(2)$$

The first term in the right hand side takes care of the explicit  time 
dependence,whereas  the  second  term  indicates  the  implicit   time 
dependence through some time dependent parameters $q$.The  parameters $q$ 
can even be identified as the co-ordinates  of  the  projective  space 
$\cal P$[4,7].The adiabatic parameters of Berry[1] belongs,in particular,to a 
subclass of this general set of parameters.In the following we use the 
same notation $\cal P$ to denote either of these two spaces,the projective or 
the general parameter space, which one is  appropriate.  Nevertheless, 
our discussion is independent of  any  specific  choice  of  the  time 
dependent parameter space,  the  nonadiabatic  geometric  phase  being 
solely an intrinsic property of the projective space.

     Projecting out the the dynamical phase of the state  $|\psi>$,  eq.(1) 
reduces to

$$
{i{D \over dt} \ |\phi> = \tilde H \ |\phi>, \hspace{.5cm}  
\tilde H = H - E} \eqno(3)$$

\noindent where $|\phi>=e^{i\int E \ dt}|\psi>$.The  state  $|\phi>$  now  belongs  to  the  horizontal 
subspace[1] of the tangent space of $H$ at the point $|\psi>$,defined via the 
parallel transport law
$${<\phi|{D \over dt}|\phi> = 0} \eqno (4)$$

\noindent  The geometric phase $\gamma$ for a closed path $C$ in $\cal P$ 
is  now  given  by  the 
holonomy integral for the connection 1-form $A$ 
$$
{A = -<\tilde {\phi}|id{\tilde {\phi}>}, \hspace{.5cm} 
\phi = e^{i\gamma} {\tilde {\phi}}, \hspace{.5cm} \tilde {\phi}\in  \cal P}
\eqno(5)$$

\noindent $d$  being  the  exterior  derivative  in $\cal P$.For  an   open   path   the 
corresponding phase (the Pancharatnam phase) is given by  an  integral 
along the shortest geodesic joining the initial and the final rays[8].
     
     Alternatively,   eq.(3)   offers   itself   to   a    nonstandard 
interpretation[3]: one could, in fact, define an intrinsic concept  of 
time in the dynamics using the nontriviality of  the  connection  form 
(5). To begin with,recall that  there  is  an  alternative  derivation 
which  realizes  the  anholonomic  phase $\gamma$  instead  as  a  dynamical 
phase[4].Introduce a unitary transformation on the state $|\phi>$
$$
{|\phi> = U \ |\chi>}\eqno(6)$$

\noindent so that the transformed state $|\chi>$  may  only  have 
 an  implicit  time 
dependence: ${\partial \over {\partial}t}|\chi>=0$.Eq.(3) becomes
$$
{i{d \over dt} \ |\chi> = h_0 \ |\chi>}\eqno (7)$$
\noindent where $h_0 = U^{\dagger} {\tilde H} U  -  i{U^{\dagger}}{\dot U}$,
and $\dot U={\partial U \over {\partial}t}$.We  note  that  such  a  unitary 
transformation always exists[4](see Appendix for a proof).Eqns.(6)-(7) 
and (4) now give $i<\chi|id\chi> = <\chi|h_0|\chi>dt =
 -i\chi|U^{\dagger}\partial U|\chi> = -i<\phi|{\partial}\phi> = A$, when 
one makes use of $\partial U|\chi>=\partial(U|\chi>)$. 
Thus the  phase  $\gamma$  is  realized  as  a 
dynamical phase for eq.(7).Note that the parameter $t$ in  the  implicit 
time derivative in the lhs of eq.(7) parametrizes the path  traced  by 
the  quantal  state  as  an  integral  curve  of  the   vector   field 
$(d/dt)=(dq_i /dt)(\partial/{\partial q_i})$  (cf.eq.(2))  in $\cal P$ 
 and  acts   as   a   dummy 
variable.Note also the reparametrization invariance of the equality 
$$
{<\chi|id\chi> = -i<\chi|{U^\dagger}{\partial U}|\chi>} \eqno(8)$$

\noindent which lies at the heart of our discussion of the internal time.
     
     The philosophy  of  the  internal  time, advocated  originally  by 
Leibniz, was discussed at lenght by Barbour[5] recently. This  intrinsic 
Leibniz concept of time (and also space) satisfies  the  Mach-Einstein 
relativity principle. As pointed  out  by  Barbour  a  gauge  invariant 
formalism of the dynamics using intrinsic (relative) concepts ought to 
yield nontrivial predictions. The inherent reparametrization invariance 
in the intrinsic  description, for  instance, results  in  a  nontrivial 
Hamiltonian  constraint  even  in  the   ordinary   (nongravitational) 
dynamics. The internal time introduced on  the  basis  of  eq.(8)  (and 
eq.(7)) (see below) clearly is a realization of the so called  Leibniz 
time.This  offers  a  most  economical  description  of  the   quantal 
evolution of  the  fluctuating  state  over  the  instantaneous  state 
$|\phi>$, without necessitating a reference to the external  Newtonian  time 
$t$.For,the constraint  alluded  to  above  corresponds,in  the  present 
discussion,to $<\phi|H-E|\phi>=0$ ,which in the adiabatic limit  leads 
 to  the 
generalized WD equation[3]
$$
{\tilde H|\phi> \equiv (H-E) |\phi> =0}\eqno(9)$$
for a fixed $E$.Recall  that  the  intrinsic  geometric  motion  in  the 
instantaneous eigenstate $|\phi>$ is encoded in the parallel transport  law 
(4).Eq.(8),on  the  otherhand,  treats  this  geometric  motion  on  a 
'dynamical' footing by transferring to a quantal moving  frame  eq.(6) 
attached to $|\phi>$.The phase $\gamma$ now corresponds to the 
correction (due  to 
quantal fluctuations) $\Delta E$ over the energy $E$,provided 
the operator $U$  is 
identified with the interaction picture evolution operator:
$$
{i{\partial \over {\partial}{\tau}} \ U = \tilde H U(\tau), \hspace{.5cm}
\tilde H(\tau) = H(\tau) - H_0}\eqno(10)$$              
\noindent where $H_0 =H(t_0 )=H(\tau = 0)$ and $U(0)=I$.
Following  the  Leibniz  view[5], the 
internal time $\tau$   must truly be an intrinsic variable, which  
relates  $d\tau$   
uniquely to the Fubini-Study arclength giving the distance between two 
infinitesimally separated states in the projective  space  $\cal P$[3].  This
follows from the fact that the (dimensionless) Fubini-Study  arclength 
$ds$(say) relates to the actual  motion  of  the  nonstationary  quantal 
state $|\psi>$ in the  projective  space $\cal P$  through  
the  gauge  invariant relation $ds=2{\Delta E}{d\tau}$ [7].\footnote{
In ref.[7] the relation $ds=2{\Delta}E{dt}$  is  obtained  in  the  ordinary 
quantum mechanics. However, $ds$ is a  property  of  the  unparametrized 
curve in $\cal P$ and hence the parameter $t$ acts as a dummy  variable.
In  the 
present intrinsic description the dummy variable $\tau$  is  raised  to  the 
status of the (internal) time via the definition $d{\tau} =ds/2{\Delta}E$.}
 The internal time $\tau$ 
 is thus  uniquely  defined 
and  is  independent  of  the  choice  of  the   parameters   $q$   .The 
distinguished intrinsic variable ($\tau$) thus makes explicit the dynamical 
content  of  the  parallel  transport  law   (4)   by   breaking   its 
reparametrization invariance dynamically.
  
   The intrinsic Schrodinger equation  which  follows  the  residual 
geometric motion of the instantaneous eigenstate $|\phi>$ is thus  obtained 
as

$$
{i{d \over d{\tau}} \ |\chi(\tau)> = -h(\tau) \ |\chi(\tau)>}\eqno(11)$$

\noindent where $ h(\tau)  =  U^{\dagger}(\tau){\tilde H}(\tau)U(\tau)$. 
Note  also   that  $<\chi| h|\chi>= \sqrt{<\psi|{\tilde H}^2|\psi>}$, by 
construction.\footnote{
 It    follows    from    eqs.(2)-(5)    and     (11)     that 
$-iU{d\over{d{\tau}}}U^{\dagger} =i{D\over{dt}}=i{\partial\over{{\partial}t}}   
+{\tilde H}+{{\tilde H}^2 \over {\Delta}E}$.The implicit time derivative 
in  eq.(1)  is 
thus related to energy uncertainty, in contrast to the concomitance of 
the explicit derivative and energy.}
Eqs.(10)  and  (11)  are  the  exact  analogs  of   an 
equivalent set of equations viz.,eqs.(8) and  (9)  in  Ref.[3,the  CQG 
paper] for the  heavy  (gravitational)  degrees  of  freedom  and  the 
lighter matter states respectively,which were derived from  a  quantum 
cosmological WD equation by using the  semiclassical  Born-Oppenheimer 
approximation.The  present  treatment,  on  the  otherhand,  is  fully 
quantal and of more general nature.It is gratifying that the  internal 
time   introduced here turns out to be consistent  with  the  relevant 
concept in quantum gravity(cosmology).An 'internal observer' belonging 
to the 'WD state' $|\phi>$  will  then  identify  the  state  $|\chi(\tau)>$
  as  a 
vertical state indicating the (small scale) evolution  in  time $\tau$  .The 
internal  state  $|\chi(\tau)>$  and  the  external  horizontal  
state   $|\phi(t)>$ 
(cf.eq.(4)) are,however, related by a gauge  rotation.We  remark  that 
eq.(11) offers novel  physical  predictions.In  fact,in  an  intrinsic 
description based on Eq.(11) one not only identifies  the  uncertainty 
$\Delta E$ in the state $|\psi>$ (eq.(1)) as the energy  for  
the  (internal)  state 
$|\chi>$, but  the  original  energy  $E$  of  $|\psi>$ 
 turns  out  also   to   be 
unobservable[3].This should be contrasted with the relative  smallness 
of $\Delta E$ (actually unobservable, in the adiabatic limit) in the extrinsic 
description,eq.(1).

     What if one  makes  use  of  the  original  external  time  $t$  in 
eqns.(10)-(11) instead of $\tau$ ?Clearly,in this extrinsic description time 
$t$ scales as $E^{-1}$. The fluctuations  in  the  state  would  therefore  be 
small,at  least  in  the  adiabatic  limit($t \rightarrow t_0$).  
Then  $\Delta E \rightarrow 0$  as 
$\tilde t \rightarrow 0, \tilde t=t-t_0$. Eq.(11),on the  otherhand,  
indicates  that  $\Delta E$  scales  as 
$-(t-t_0)^{-1}$. This dichotomy can be resolved only if one sets the  initial 
time  uniquely at $t_0 = -{\infty}$ .This  choice  of  initial  condition,  though 
appears reasonable for an isolated  scattering  state,is  nevertheless 
too restrictive for strongly interacting systems.The present intrinsic 
description should therefore  be  of  interest  not  only  in  quantum 
gravity but also in posing the initial value  problem  for  a  general 
(strongly) interacting quantum system.

     Note that the (-)sign in eq.(11) takes care of the fact that  the 
direction of  traversal of a path in $\cal P$ as seen externally is  reversed 
internally.Further,as noted above the internal time scales as  
$(\Delta E)^{-1}$,
 in contrast to the external time behaviour:
$\Delta E \rightarrow 0$  as  $\tilde t \rightarrow 0$.Moreover,  in 
view of the relation $t \approx E^{-1}$; $\tilde t \approx E^{-2}{\delta E}$
 and thus the $\tilde t \rightarrow 0$ limit  is  reached 
either as $\delta E \rightarrow 0$ (the external stationarity) or 
by taking $E \rightarrow \infty$ (asymptotic 
high energy region). Further,in relation to a  very  small  time  scale 
(when the concept of the internal  time  becomes  relevant)  both  the 
energy differential $\delta E$ and the uncertainty $\Delta E$ will  
appear  large  and 
comparable, $\approx O(E)$. In that limit of large fluctuations 
time variables  $\tilde t$ 
and $\tau$  are related by the duality relation $\tilde t=1/{\tau}$. 
Here,we  measure  both 
the time variables in the unit of $E$. In the following we  omit  tilde 
and assume $t \rightarrow 0$. We next show by following the  RG  
technique  that  the 
behaviour of a quantal evolution may undergo a change at the self-dual 
point  $t=\tau(=1)$, analogous  to  a  phase  transition   at   a   critical 
point.However, before delving into this discussion we  first  give  an 
example illustrating briefly some of the results discussed so far.

     Let us consider a time dependent harmonic oscillator given by the 
Hamiltonian[9], $H(t)=(p^2/2m) + mw^2(t)q^2/2$. The time dependent 
state  $|\chi>$ can 
be determined by the  so(2,1)  valued  invariant  operator  method[for 
details we refer to [9]].The mean energy in the state $|\psi>$,in the short 
time limit $t \rightarrow 0$ near an initial adiabatic state  $|\psi(0)>$
 can  be  written 
as $ E=E_0 (1+\nu)$,  where $E_0 =w_0(n+1/2)$, $w_0 =w(0)$, $\nu=2|\beta|^2$,
 $\beta$=the Bogoliubov 
$\beta$-coefficient. Then $\Delta E = {\nu}E_0 \approx$the uncertainty 
in  $|\psi>$  when $\nu \approx O(1)$.Further 
the nonadiabatic phase can be computed[10,3] as $\gamma={\Delta E}t$. 
Note  that  the 
total phase remains equal to  $(E_0t)$,  in  the  present  approximation. 
Clearly as the external time scales  as  $t \approx {\nu}E$, 
$\nu \rightarrow 0$; ${\Delta E}t \approx {\nu}^2 \rightarrow 0$. 
However, $\gamma$ 
can be freezed in the time variable  $\tau \approx ({\nu}E)^{-1}$, 
letting  ${\Delta E}{\tau} \approx 1$. In  this 
intrinsic description  $\tau \rightarrow 0  \Rightarrow \nu \rightarrow
\infty$. Interestingly, the  scaled  Hamiltonian 
in Eq.(11) can be obtained  as  $h(\tau) = {\nu}H(\tau)$. Further  details, 
however, 
will be reported in ref.[10].We  conclude  this  illustration  with  a 
remark.The adiabatic initial condition is assumed here for the sake of 
computational ease.In  principle  one  could  as  well  start  with  a 
strongly interacting initial  state,  which  would  then  lead  to  an 
adiabatic state in the dual intrinsic description. 

     To return to the main body of the analysis,let us recall that our 
intention is to study the nature of the  quantal  evolution  near  the 
self dual point $t=\tau$. To this effect,let ${\psi}_{RG}(0,t)$ 
denote the  two  point 
time-time  correlation  function  of   the   (Euclidean)   Schrodinger 
equation. Then ${\psi}_{RG} = <\psi(0)|\psi(t)>$, where   $|\psi>$  
denotes  the  purely  time 
dependent  factor  of  the   Euclidean   Schrodinger   wave   function 
henceforth.One may interpret it as the transition probability  of  the 
state from the initial time $t=0$ to the final time $t$, with  energy  $E$. 
In the  conventional  extrinsic  treatment   of   quantum   mechanics, the 
probability is 1, in  the  adiabatic  limit  $t \rightarrow 0$. 
In  reality,  however, 
fluctuations are nonzero, making room for an intrinsic description  as 
well. A given state may thus have two possible paths, given by  either 
of the two  equations  viz.,eq.(1)  or  (11),for  its  evolution  near 
$t=0(\tau =0)$.Consequently,as we now show, the  probability  ceases  to  be 
unity.
     
     Writing  ${d\gamma \over dt} = {\nu}E$, and introducing  the  Wick 
 rotation $t \rightarrow -it$, eq.(1) 
(together with (2)) is expressed in the form of a  Callan-Symanzik  RG 
equation[11]

$$
{({1 \over E}{\partial \over {\partial}t} + \beta(\tau)
{\partial \over {\partial}{\tau}} + 1 + \nu ) \psi_{RG} = 0}
\eqno (12)$$

\noindent where the $\beta$-function is given  by 
$\beta(\tau)={d{\tau} \over dt}$,  and  $Ed\tau \rightarrow d\tau$. 
Eq.(12)  is 
obtained by projecting the Wick rotated equation (1)  on  the  initial 
state  $|\psi(0)>$, since  ${D \over dt} <\psi(0)|\psi(t)>
=<\psi(0)|{D \over dt}\psi(t)>$.  The   homogeneous 
equation   (12)   restricts   $t>0$. Note   that   in    the    extrinsic 
description, $\tau \equiv t$;  leading  to  the  trivial  scaling  
(see   below). We 
are, however, interested in the nontrivial case when $\tau$  is the  internal 
time. The $\beta$-function is then computed by recalling that  $t$  
and $\tau$    are 
duality variables: $\beta(\tau) =-\tau^2$. $\tau = 0$  is  thus  
a(n)  (ultraviolate  stable) 
fixed point of the $\beta$-function. This could be interpreted as a  sort  of 
an 'asymptotic freedom' in the nonstationary dynamical system. In  fact 
the  nontrivial  'asymptotic  freedom'  is  a   manifestation   of   a 
competition  between  two   different   limits   $t \rightarrow 0$ 
(externally)   and 
$\tau \rightarrow \infty$ (internally)   near   the   fixed   point. 
Consequently, a   strongly 
interacting system could be mapped to a weakly interacting one by  the 
duality transformation. Another important effect of this duality is the 
stabilization of the 'anomalous  dimension' $\nu$  to  a  nonzero  finite 
value. This is obtained by comparing the two Schrodinger equations  (1) 
and  (11), which  should  agree  at   the   fixed   point. One   obtains 
$(1+\nu)Edt = -{\nu}Ed{\tau} \Rightarrow 1+\nu ={\nu}{\tau}^2$, 
which in the limit  $t \rightarrow 0$, $\tau \approx \nu^{-1} \rightarrow
\infty$;  gives  the 
fixed point equation  $1+\nu_0 =\nu_0^{-1}$, yielding  the 
 limiting  value  of  the 
'anomalous dimension' $\nu$ as the golden mean $\nu_0 = 
{1 \over 2}(\sqrt5 -1)$. The nonzero  value 
of $\nu$ in the present case should be contrasted with  the  vanishing  of 
the same in the case of a gauge theory  asymptotic  freedom. The  final 
form of the solution of the RG equation (12) can  now  be  written  as 
$\psi_{RG}  = \sigma^{-1-\nu}g(\tau )$, where  $g$  is  an  arbitrary 
 function  of  the  running 
'coupling constant' $\tilde {\tau}$  satisfying the boundary condition  
$g$=const.,  at 
 $\tilde {\tau}=0$.. Here  $\sigma$    is  an  energy  scaling  variable. 
One  thus  obtains  the 
nontrivial scaling for the  two  point  correlation  function  in  the 
asymptotic region : $\psi_{RG} \approx \sigma^{-1-\nu_0}$. 
For the sake of comparison, we note that 
the expected scaling  in  the  extrinsic  description,  which  can  be 
directly verified is $\psi^0_{RG} \approx \sigma^{-1}$. 
To get better insights into  the  scaling 
law,we take the inverse Laplace transform(as we consider $t>0$)  of  the 
same so as to express it in the  time coordinate: 
$\psi_{RG} \approx t^{\nu_0} (\psi^0_{RG} \approx 1)$.

     Clearly, a natural interpretation of this scaling behaviour could 
be obtained in the framework of the fractal geometry.The  scaling  law 
tells us that the fixed point $t=0$ is an  extended  object  with  finer 
fractal-like structures.By the  time  translation  invariance  of  the 
Schrodinger equation it follows that every point in the time axis  has 
identical fractal structure.The extended  time  axis  is  thus  a {\it  fat 
fractal} [12],with the uncertainty exponent $\nu_0$. An  interesting  physical 
implication of this uncertainty exponent  is  the  following: Near  the 
fixed point both the Schrodinger equations  (1)  and  (11)  could,  in 
principle, be considered  as  the  correct  evolution  equations  with 
definite   but   distinct   physical   predictions   for    a    given 
system.Moreover,these  distinct  physical  possibilities  of  a  given 
state,are in a state of an inextricable mixture in view of the fractal 
nature of the fixed  point.An  experiment  intending  to  measure  the 
(geometric) phase (for a very  short  time  evolution)  of  the  state 
should ,therefore, reveal a  randomness  in  the  observed  values.One 
expects  to  see,  for  instance,a  data  consisting   of   a   random 
distribution of two numbers,e.g.,1(suitably normalized  value  of  the 
phase on the basis of eq.(1))  and  0(for  eq.(11)),provided  repeated 
measurements  are  made,  on  identically  prepared  states,   for   a 
considerable  period   of   time.A   very   high   resolution   (phase 
determination) experiment on a  nonadiabatic  quantal  system  in  the 
region of (relatively) large  fluctuations  may  perhaps  confirm  the 
prediction.If so,this may be considered as an  experimental  test  for 
the  fractal  nature  of  time,which  might  have  important  physical 
implications[3].We conclude by noting that  by  inhabiting  a  duality 
transformation,the fractal time extends the  linear  time  translation 
invariance of the Schrodinger equation to an  action  of  the  SL(2,R) 
transformations.Further  details  of  this  group  action  in  quantum 
mechanics and the other related issues will be considered separately.


\section*{Acknowledgements}
It is a pleasure to thank 
T.Padmanabhan  and  S.Sinha 
for  discussions  and  Inter  University  centre  for  Astronomy   and 
Astrophysics,Pune for awarding a Senior Associateship. The work is also 
supported by the Department of Science  and  Technology,Government  of 
India.

\section*{APPENDIX}

Theorem:Given the Schrodinger equations (3)  and  (7),there  exists  a 
unitary transformation $U$ (Eq.(6)) between them provided $({\tilde H}-h )$ 
is  not implicitly time dependent .
     
     The proof follows from an explicit construction.
Let $U=U_0(t,t_0 )U_{1}^{\dagger}(t,t_0)$, where $U_0$  and $U_1$  
 be the evolution operators  of 
Eqs.(3) and (7) respectively. Then $U$  satisfies  the  relation $h_0 
=U^{\dagger}{\tilde H}U -iU^{\dagger}{\dot U}$, provided $h_0 
=iU_{0}^{\dagger} {d\over dt}U_0$.. The theorem  follows  once  we  split  
$U_0$  further into an explicit and an implicit time dependent unitary opera- 
tors:$U_0 =U_{ex}U_{im}$. This splitting is certainly possible 
for the class of 
models with $H=H_{ex}+ H_{im}$, $[H_{ex},H_{im}]=0$.
In the present discussion $H_{ex}=0$.

\end{document}